\documentclass[twocolumn,aps,prl]{revtex4}

\usepackage{graphicx}
\usepackage{amsmath}

\graphicspath{{D:/Yale/Papers/ACStark/preliminary_versions/figures/}}

\newcommand{\eq}[1]{(\ref{#1})}

\begin{document}

\title{AC-Stark Shift and Dephasing
of a Superconducting Qubit\\ Strongly Coupled to a Cavity Field
}

\author{D.~I.~Schuster}
\affiliation{Departments of Applied Physics and Physics, Yale
University, New Haven, CT 06520}
\author{A.~Wallraff}
\affiliation{Departments of Applied Physics and Physics, Yale
University, New Haven, CT 06520}
\author{A.~Blais}
\affiliation{Departments of Applied Physics and Physics, Yale
University, New Haven, CT 06520}
\author{L.~Frunzio}
\affiliation{Departments of Applied Physics and Physics, Yale
University, New Haven, CT 06520}
\author{R.-S.~Huang}
\altaffiliation{Department of Physics, Indiana University,
Bloomington, IN 47405} \affiliation{Departments of Applied Physics
and Physics, Yale University, New Haven, CT 06520}
\author{J.~Majer}
\affiliation{Departments of Applied Physics and Physics, Yale
University, New Haven, CT 06520}
\author{S.~M.~Girvin}
\affiliation{Departments of Applied Physics and Physics, Yale
University, New Haven, CT 06520}
\author{R.~J.~Schoelkopf}
\affiliation{Departments of Applied Physics and Physics, Yale
University, New Haven, CT 06520}
\date{\today, submitted to \emph{PRL}}

\begin{abstract}
We have spectroscopically measured the energy level separation of
a superconducting charge qubit coupled non-resonantly to a single
mode of the electromagnetic field of a superconducting on-chip
resonator. The strong coupling leads to large shifts in the energy
levels of both the qubit and the resonator in this circuit quantum
electrodynamics system. The dispersive shift of the resonator
frequency is used to non-destructively determine the qubit state
and to map out the dependence of its energy levels on the bias
parameters. The measurement induces an ac-Stark shift of $0.6 \,
\rm{MHz}$ per photon in the qubit level separation. Fluctuations
in the photon number (shot noise) induce level fluctuations in the
qubit leading to dephasing  which is the characteristic
back-action of the measurement. A cross-over from lorentzian to
gaussian line shape with increasing measurement power is observed
and theoretically explained. For weak measurement a long intrinsic
dephasing time of $T_2 > 200 \, \rm{ns}$ of the qubit is found.
\end{abstract}

\maketitle

We have recently demonstrated that a superconducting quantum
two-level system can be strongly coupled to a single microwave
photon \cite{wallraff04c}. The strong coupling between a quantum
solid state circuit and an individual photon, analogous to atomic
cavity quantum electrodynamics (CQED) \cite{Mabuchi02}, has
previously been envisaged by many authors, see
Ref.~\onlinecite{Blais04} and references therein. Our circuit
quantum electrodynamics architecture \cite{Blais04}, in which a
superconducting charge qubit, the Cooper pair box
\cite{Bouchiat98}, is coupled strongly to a coplanar transmission
line resonator, has great prospects both for performing quantum
optics experiments \cite{Walls94} in solids and for realizing
elements for quantum information processing \cite{Nielsen00} with
superconducting circuits \cite{scqc}.

In this letter we present spectroscopic measurements which
demonstrate the \emph{non-resonant (dispersive)} strong coupling
between a Cooper pair box and a coherent microwave field in a high
quality cavity. The quantum state of the Cooper pair box is
controlled using resonant microwave radiation and is read out with
a dispersive quantum non-demolition (QND) measurement
\cite{Grangier98,Blais04,Nogues99}. The interaction between the
Cooper pair box and the measurement field containing $n$ photons
on average gives rise to a large ac-Stark shift of the qubit
energy levels, analogous to the one observed in CQED
\cite{Brune94}. As a consequence of the strong coupling, quantum
fluctuations in $n$ induce a broadening of the transition line
width, characterizing the back action of the measurement on the
qubit.

\begin{figure}[tbp]
\includegraphics[width = 0.8\columnwidth]{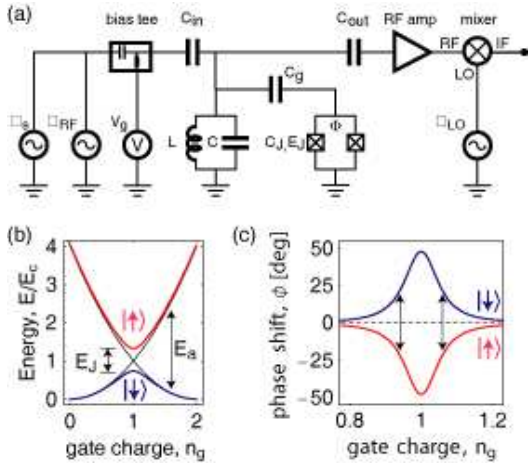}
\caption{(color online) (a) Simplified circuit diagram of
measurement setup. The phase $\phi$ and amplitude $T$ of a
microwave at $\omega_{\rm{RF}}$ transmitted through the resonator,
amplified and mixed down to an intermediate frequency
$\omega_{\rm{IF}} = \omega_{\rm{RF}}-\omega_{\rm{LO}}$ using a
local oscillator at $\omega_{\rm{LO}}$ is measured. An additional
spectroscopy microwave at $\omega_{\rm{s}}$ is applied to the
input port of the resonator. (b) Ground
$\left|\downarrow\right\rangle$ and excited
$\left|\uparrow\right\rangle$ state energy levels of CPB
\textit{vs.} gate charge $n_g$. (c) Calculated phase shift $\phi$
in ground and exited state \textit{vs.} $n_g$ for
$\Delta_{\rm{a,r}}/2\pi = 100 \, \rm{MHz}$.} \label{fig:setup}
\end{figure}

In our circuit QED architecture \cite{Blais04}, see
Fig.~\ref{fig:setup}a, a split Cooper pair box \cite{Bouchiat98},
modelled by the two-level hamiltonian $H_a = - 1/2
\left(E_{\rm{el}} \, \sigma_x + E_{\rm{J}} \, \sigma_z\right)$
\cite{Makhlin01}, is coupled capacitively to the electromagnetic
field of a full wave ($l = \lambda$) transmission line resonator,
described by a harmonic oscillator hamiltonian $H_{\rm{r}} = \hbar
\omega_{\rm{r}} (a^{\dagger}a + 1/2)$. In the Cooper pair box, the
energy difference $E_a = \hbar \omega_{\rm{a}} =
\sqrt{E_{\rm{el}}^2+E_{\rm{J}}^2}$ between the ground state
$\left|\downarrow\right\rangle$ and the first excited state
$\left|\uparrow\right\rangle$, see Fig.~\ref{fig:setup}b, is
determined by its electrostatic energy $E_{\rm{el}} = 4 E_{\rm{C}}
(1-n_{\rm{g}})$ and its Josephson coupling energy $E_{\rm{J}} =
E_{\rm{J,max}} \cos\left(\pi \Phi_{\rm{b}}\right)$. Here,
$E_{\rm{C}} = e^2/2C_{\rm{\Sigma}} \approx 5.0 \, \rm{GHz}$ is the
charging energy given by the total box capacitance $C_{\Sigma}$,
$n_{\rm{g}} = C_g^{\star}V_{\rm{g}}/e$ is the gate charge
controlled by the gate voltage $V_g$ applied through a gate with
effective capacitance $C_g^{\star}$, and $E_{\rm{J,max}} \approx
8.0 \, \rm{GHz}$ is the maximum Josephson coupling energy of the
two junctions which is modulated by applying a flux bias
$\Phi_{\rm{b}} = \Phi/\Phi_0$ to the loop of the split box, see
Fig.~\ref{fig:setup}a. Near its resonance frequency $\omega_r =
1/\sqrt{LC} \approx 2\pi \, 6.0 \, \rm{GHz}$, the resonator is
accurately modelled as a harmonic oscillator with lumped
inductance $L$ and capacitance $C$.

In the presence of strong mutual coupling between the qubit and
the resonator \cite{wallraff04c}, their \emph{dressed} excitation
energies $\widetilde{\omega}_{\rm{a}}$ and
$\widetilde{\omega}_{\rm{r}}$, are modified from their bare values
$\omega_{\rm{a}}$ and $\omega_{\rm{r}}$. For large detuning
$\Delta_{\rm{a,r}} = \omega_{\rm{a}} - \omega_{\rm{r}}$ the
dressed energy levels are determined by the Hamiltonian
\cite{Blais04}
\begin{equation}\label{eq:dispersiveHalitonianResonator}
    H \approx \hbar \left(\omega_{\rm{r}} +
\frac{g^2}{\Delta_{\rm{a,r}}}\sigma_z\right)a^{\dagger}a +
\frac{1}{2}\hbar
\left(\omega_{\rm{a}}+\frac{g^2}{\Delta_{\rm{a,r}}}\right)\sigma_z
\, ,
\end{equation}
where $g/2\pi \approx 5.8 \, \rm{MHz}$ is the coupling strength
between a single photon and the qubit \cite{wallraff04c}. In this
non-resonant case, the dressed resonator frequency
$\widetilde{\omega}_{\rm{r}} = \omega_{\rm{r}} \pm
{g^2}/{\Delta_{\rm{a,r}}}$ depends on the qubit state $\sigma_z =
\pm 1$ and the detuning $\Delta_{\rm{a,r}}$. The qubit state can
thus be inferred from the phase shift $\phi$ that a probe
microwave transmitted through the resonator at frequency
$\omega_{\rm{RF}}$ experiences due to the interaction with the
qubit \cite{wallraff04c,Blais04}. In Fig.~\ref{fig:setup}c, the
phase shift $\phi = \pm
\tan^{-1}\left(2g^2/\kappa\Delta_{\rm{a,r}}\right)$, where $\kappa
= \omega_{\rm{r}}/Q$ is the decay rate of photons from the
resonator with quality factor $Q \approx 10^4$, is plotted versus
gate charge $n_{\rm{g}}$.  $\phi$ is maximum at $n_{\rm{g}}=1$
where the detuning $\Delta_{\rm{a,r}}$ is smallest and falls off
as the detuning is increased with increasing $n_g$. Moreover,
$\phi$ has opposite signs in the ground
$\left|\downarrow\right\rangle$ and excited
$\left|\uparrow\right\rangle$ states of the CPB.

\begin{figure}[tbp]
\includegraphics[width = 1.0 \columnwidth]{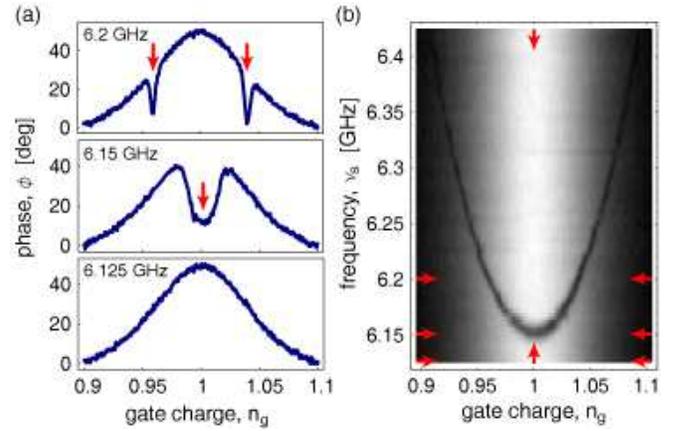}
\caption{(color online) (a) Probe microwave phase shift $\phi$
\textit{vs.} gate charge $n_{\rm{g}}$ at spectroscopy frequency
$\nu_{\rm{s}} = 6.125 \, \rm{GHz}$ (lower panel), $6.15 \,
\rm{GHz}$ (middle panel) and $6.2 \, \rm{GHz}$ (upper panel). (b)
Density plot of $\phi$ \textit{vs.}~$n_{\rm{g}}$ and
$\nu_{\rm{s}}$, white (black) corresponds to large (small) phase
shift. Horizontal arrows indicate line cuts shown in (a), vertical
arrows indicate line cuts shown in
Fig.~\ref{fig:MeasurementLineWidthCombi}a. }
\label{fig:SpectroscopyCombi}
\end{figure}

Qubit state transitions can be driven by applying an additional
microwave of frequency $\omega_{\rm{s}}$ and power $P_{\rm{s}}$
resonant with the qubit ($\Delta_{\rm{s,a}} =
\omega_{\rm{s}}-\widetilde{\omega}_{\rm{a}} = 0$) to the input
port of the resonator, see Fig.~\ref{fig:setup}a. On resonance
($\Delta_{\rm{s,a}} = 0$) and for a continuous (cw) large
amplitude spectroscopy drive, the qubit transition saturates and
the population in the exited and ground state approaches $1/2$. In
this case, the measured phase shift of the probe beam at
$\omega_{\rm{RF}}$ is expected to saturate at $\phi = 0$, see
Fig.~\ref{fig:setup}c. By sweeping the spectroscopy frequency
$\omega_{\rm{s}}$ \textit{vs.}~the gate charge $n_{\rm{g}}$ and
\emph{continuously} measuring $\phi$ we have mapped out the energy
level separation $\widetilde{\omega}_{a}$ of the qubit, see
Fig.~\ref{fig:SpectroscopyCombi}. In the lower panel of
Fig.~\ref{fig:SpectroscopyCombi}a, the measured phase shift $\phi$
is shown for the non-resonant case, where $\omega_{\rm{s}} <
\widetilde{\omega}_{\rm{a}}$ for all values of gate charge
$n_{\rm{g}}$. The measured phase shift is, as expected, a
continuous curve similar to the one shown in
Fig.~\ref{fig:setup}c. In the middle panel of
Fig.~\ref{fig:SpectroscopyCombi}a, the spectroscopy microwave at
$\nu_{\rm{s}} = \omega_{\rm{s}}/2\pi = 6.15 \, \rm{GHz}$ is in
resonance with the qubit at $n_{\rm{g}}=1$, populating the excited
state and thus inducing a dip in the measured phase shift $\phi$
around $n_{\rm{g}}=1$, as expected. Note that, as predicted
\cite{Blais04}, our measurement scheme has the advantage of being
most sensitive at charge degeneracy, a bias point where
traditional electrometry, using a radio frequency single electron
transistor (rf-SET) \cite{Lehnert03} for example, is unable to
distinguish the qubit states.

When $\nu_{\rm{s}}$ is increased to higher values, resonance with
the qubit occurs for two values of $n_{\rm{g}}$ situated
symmetrically around $n_{\rm{g}}=1$, leading to two symmetric dips
in $\phi$, see upper panel of Fig.~\ref{fig:SpectroscopyCombi}a.
From the $[n_g,\nu_s]$ positions of the spectroscopic lines in the
measured phase $\phi$, the Josephson energy $E_{\rm{J}} = 6.2 \,
\rm{GHz}$ and the charging energy $E_{\rm{C}} = 4.8 \, \rm{GHz}$
are determined in a fit using the full qubit Hamiltonian
\cite{Makhlin01}, see density plot of $\phi$ \textit{vs.}
$n_{\rm{g}}$ and $\nu_s$ in Fig.~\ref{fig:SpectroscopyCombi}b. In
this experiment the flux bias $\Phi_{\rm{b}}$ has been chosen to
result in a minimum detuning of about $\Delta_{\rm{r,a}}/2\pi
\approx 100 \, \rm{MHz}$ at $n_{\rm{g}}=1$. The tunability of
$E_{\rm{J}}$ (i.e.~the detuning at charge degeneracy) has been
demonstrated previously \cite{wallraff04c}. It is also worth
noting, that the spectroscopy frequency $\omega_{\rm{s}}$
typically remains strongly detuned ($\Delta_{\rm{s,r}} =
\omega_{\rm{s}}-\omega_{\rm{r}} > 2 \pi \, 100 \,\rm{MHz}$) from
the resonator, such that a large fraction of the spectroscopy
photons are reflected at the input port and only a small number
$n_s$, determined by the lorentzian line shape of the resonator,
populates the resonator.

Various other radio or microwave frequency qubit read-out schemes
have been developed recently \cite{Lehnert03,Lupascu03,Siddiqi03}.
In particular in a closely related experiment, the level
separation of a split Cooper pair box coupled \emph{inductively}
to a \emph{low frequency, moderate $Q$} tank circuit has been
determined spectroscopically \cite{Born03}.

\begin{figure}[tbp]
\includegraphics[width = 0.85 \columnwidth]{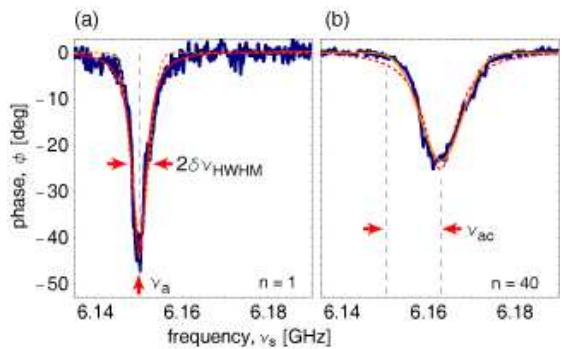}
\caption{(color online) Measured spectroscopic lines (blue lines)
at (a) intra resonator photon number $n \approx 1$ ($P_{\rm{RF}} =
-30 \, \rm{dBm}$) with fit to lorentzian line shape (solid line)
and at (b) $n \approx 40$ ($P_{\rm{RF}} = -16 \, \rm{dBm}$) with
fit to gaussian line shape (solid line). Dashed lines are best
fits to (a) gaussian or (b) lorentzian line shapes, respectively.
The qubit transition frequency $\nu_{\rm{a}}$ at low
$P_{\rm{RF}}$, the half width half max $\delta\nu_{\rm{HWHM}}$ and
the ac-Stark shift $\nu_{\rm{ac}}$ of the lines are indicated.}
\label{fig:MeasurementLineWidthCombi}
\end{figure}

\begin{figure}[!t]
\includegraphics[width = 1.0\columnwidth]{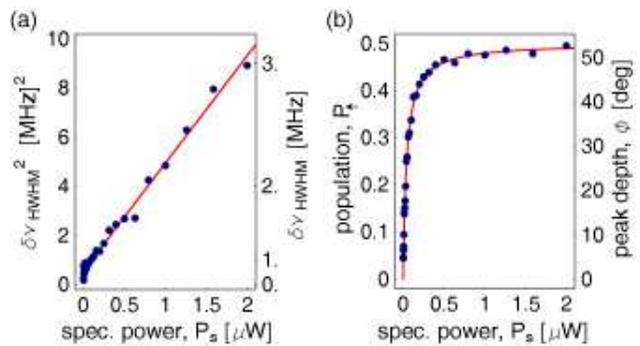}
\caption{(color online) (a) Measured qubit line width
$\delta\nu_{\rm{HWHM}}$ \textit{vs.} input spectroscopy power
$P_s$ (solid circles) with fit (solid line). Probe beam power
$P_{\rm{RF}}$ is adjusted such that $n<1$. (b) Measured peak depth
$\phi$ and excited state population probability $P_{\uparrow}$ on
resonance $\Delta_{\rm{s,01}}=0$ \textit{vs.} $P_{\rm{s}}$ (solid
circles) with fit (solid line). } \label{fig:PowerBroadeningCombi}
\end{figure}

The width and the saturation level of the spectroscopic lines
discussed above depend sensitively on the power $P_{\rm{s}}$ of
the spectroscopic drive. Both quantities can be extracted from the
excited state population
\begin{equation}\label{eq:lineshape}
        P_{\uparrow} = 1 - P_{\downarrow} =
        \frac{1}{2}\frac{n_s\omega_{\rm{vac}}^2 T_1 T_2}{1+\left(T_2
        \Delta_{\rm{s,01}}\right)^2+n_s\omega_{\rm{vac}}^2 T_1 T_2}\, ,
\end{equation}
found from the Bloch equations in steady state \cite{Abragam61},
where $\omega_{\rm{vac}} = 2 g$ is the vacuum Rabi frequency,
$n_s$ the average number of spectroscopy photons in the resonator,
$T_1$ the relaxation time and $T_2$ the dephasing time of the
qubit. We have extracted the transition line width and saturation
from spectroscopy frequency scans for different drive powers
$P_{\rm{s}}$ with the qubit biased at charge degeneracy
($n_{\rm{g}}=1$). We observe that the spectroscopic lines have a
lorentzian line shape, see
Fig.~\ref{fig:MeasurementLineWidthCombi}a, with width and depth in
accordance with Eq.~(\ref{eq:lineshape}).
The half width at half max (HWHM) of the line is shown to follow
the expected power dependence $2\pi\delta\nu_{\rm{HWHM}} =
{1}/{T'_2} =  \sqrt{{1}/{T_2^2} + n_{\rm{s}} \omega_{\rm{vac}}^2
{T_1}/{T_2}}$ \cite{Abragam61}, where the input microwave power
$P_{\rm{s}}$ is proportional to $n_{\rm{s}}\omega_{\rm{vac}}^2$,
see Fig.~\ref{fig:PowerBroadeningCombi}a. In the low power limit
($n_{\rm{s}}\omega_{\rm{vac}}^2 \rightarrow 0$), the unbroadened
line width is found to be small $\delta\nu_{\rm{HWHM}} \approx 750
\, \rm{kHz}$, corresponding to a long dephasing time of $T_2 > 200
\, \rm{ns}$ at $n_{\rm{g}}=1$, where the qubit is maximally
protected from 1/f charge fluctuations. At larger drive, the width
increases proportionally to the drive amplitude. The depth of the
spectroscopic dip at resonance ($\Delta_{\rm{s,a}}=0$) reflects
the probability of the qubit to be in the excited state
$P_{\uparrow}$ and depends on $P_{s}$ as predicted by
Eq.~(\ref{eq:lineshape}), see
Fig.~\ref{fig:PowerBroadeningCombi}b. At low drive the population
increases linearly with $P_{\rm{s}}$ and then approaches $0.5$ for
large $P_{\rm{s}}$.

\begin{figure}[!tp]
\includegraphics[width = 1.0\columnwidth]{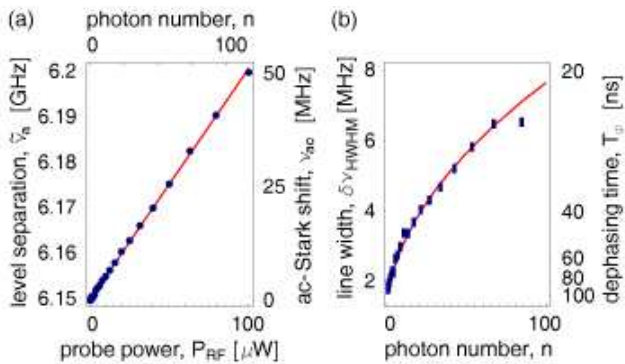}
\caption{(color online) (a) Measured qubit level separation
$\widetilde{\nu}_{\rm{a}}$ and fit (solid line) \textit{vs.}~input
microwave probe power $P_{\rm{RF}}$. The ac-stark shift
$\nu_{\rm{ac}}$ and the intra-resonator photon number $n$
extracted from the fit are also indicated. (b) Measurement
broadened qubit line width $\delta\nu_{\rm{HWHM}}$
\textit{vs.}~$n$.  The corresponding total dephasing time
$T_{\varphi} = 1/2\pi\delta\nu_{\rm{HWHM}}$ as reduced by the
measurement is also indicated. The fit (solid line) is obtained
from Eq.~\eq{eq:measurementlineshape} with a spectroscopy power
broadened $T'_2 \approx 105 \, \rm{ns}$. }
\label{fig:acStarkShiftLineWidthCombi}
\end{figure}

In the above we have demonstrated that the strong coupling of the
qubit to the radiation field modifies the resonator transition
frequency in a way that can be exploited to measure the qubit
state. Correspondingly, the resonator acts back onto the qubit
through their mutual strong coupling. Regrouping the terms of the
hamiltonian in Eq.~(\ref{eq:dispersiveHalitonianResonator}) one
sees that the \emph{dressed} qubit level separation is given by
$\widetilde{\omega}_{a} =
\omega_a+{2ng^2}/{\Delta_{\rm{a,r}}}+{g^2}/{\Delta_{\rm{a,r}}}$,
where we note that the resonator gives rise to an ac-Stark shift
of $\pm n g^2/\Delta_{\rm{a,r}}$, proportional to the
intra-resonator photon number $n = \langle a^{\dagger}a \rangle$,
as well as a Lamb shift $\pm g^2/2\Delta_{\rm{a,r}}$, due to the
coupling to the vacuum fluctuations, of the qubit levels. The
ac-Stark shift is observed in a measurement in which we perform
spectroscopy at $n_{\rm{g}}=1$ for fixed spectroscopy power
$P_{\rm{s}}$ and vary the probe beam power $P_{\rm{RF}}$, to
change the average number $n$ of measurement photons in the
resonator, see Fig.~\ref{fig:acStarkShiftLineWidthCombi}a.
As expected, it is observed that the qubit level separation
$\widetilde{\nu}_{\rm{a}} = \widetilde{\omega}_{\rm{a}}/2\pi$ is
linear in $P_{\rm{RF}}$, i.e.~the ac-Stark shift $\nu_{\rm{ac}} =
2 n g^2 / 2 \pi \Delta_{\rm{a,r}}$ being linear in the photon
number $n$ is observed. In the limit of $P_{\rm{RF}} \rightarrow
0$ ($n \rightarrow 0$), the bare qubit level separation
$\omega_{\rm{a}} + g^2/\Delta_{\rm{a,r}}= 2\pi \, 6.15 \,
\rm{GHz}$ is determined, where $g^2/\Delta_{\rm{a,r}}$ is the
small Lamb shift which can not be separated from $\omega_{\rm{a}}$
in our current experiments. Knowing the coupling constant $g$ from
an independent measurement of the vacuum Rabi mode splitting
\cite{wallraff04c} and having determined $\Delta_{\rm{a,r}}$ from
spectroscopic measurements in the $n \rightarrow 0$ limit, the
dependence of the intra-resonator photon number $n$ on the input
power $P_{\rm{RF}}$ is determined from the measured ac-Stark shift
$\nu_{\rm{ac}}$. We find that an input microwave power of
$P_{\rm{RF}} = -31 \, \rm{dBm}$ corresponds to $n=1$ which is
consistent with an intended attenuation of approximately $105 \,
\rm{dB}$ in the input coaxial line. The ac-Stark shift of the
qubit at this particular detuning is a remarkable $0.6 \,
\rm{MHz}$ per photon in the cavity and is comparable to the line
width. Using this method, the intra-resonator photon number was
accurately calibrated for the vacuum Rabi mode splitting
measurements presented in Ref.~\onlinecite{wallraff04c}.

Quantum fluctuations (photon shot noise) $\delta n$ around the
average photon number $n$ of the coherent field populating the
resonator give rise to random fluctuations in the qubit transition
frequency due to the ac-Stark shift. This leads to measurement
induced dephasing, and thus, to a broadening of the qubit line
width, see Figs.~\ref{fig:MeasurementLineWidthCombi} and
\ref{fig:acStarkShiftLineWidthCombi}. This is the measurement back
action and can be understood quantitatively by considering the
relative phase $\varphi(t) = 2g^2/\Delta_{\rm{a,r}} \int^t_0 dt'
\delta n(t')$ accumulated in time between the ground and excited
state of the qubit. Following Ref.~\onlinecite{Blais04}, the
measurement-induced phase decay of the qubit is then characterized
by
\begin{equation}\label{eq:phasefluctuations}
    \langle e^{i\varphi(t)}\rangle
    =
\exp\left[-\frac{2 g^4}{\Delta^2_{\rm{a,r}}} \iint_0^t
dt_1dt_2\langle \delta n(t_1)\delta n(t_2)\rangle\right]
\end{equation}
where the fluctuations $\delta n$ are assumed to be gaussian. In
the above expression, the photon correlation function $\langle
\delta n(t) \delta n(0)\rangle = n
\exp\left({-\kappa\left|t\right|/2}\right)$ of the coherent probe
beam in the resonator is governed by the cavity decay rate
$\kappa$ and physically represents the white photon shot noise
filtered by the cavity response. The spectroscopic line shape
$S(\omega)$ is obtained from the Fourier transform of $\langle
\exp[i\varphi(t)]\rangle e^{-t/T'_2}$ where $1/T'_2$ takes into
account dephasing mechanisms independent of the measurement
\begin{equation}\label{eq:measurementlineshape}
   S(\omega)
    = \frac{1}{\pi}
\sum_{j=0}^{\infty}\frac{\left(-4\chi\right)^j}{j!} \frac{1/T'_2 +
2 \kappa \chi +
j\kappa/2}{(\omega-\tilde\omega_a)^2+\left(\frac{1}{T'_2} + 2
\kappa \chi + \frac{j\kappa}{2}\right)^2}
    \, .
\end{equation}
The form of the line shape depends on the dimensionless parameter
$\chi=n\theta_0^2$ where $\theta_0 = 2g^2/\kappa\Delta_{\rm a,r}$
is the transmission phase shift describing the strength of the
measurement. For small $\chi$ the measurement rate is slow
compared to $\kappa$ and the phase diffuses in a random walk
$\langle \varphi(t)^2\rangle \sim 4\theta_0^2n\kappa t$ leading to
a homogeneously broadened lorentzian line of HWHM of
$2\theta_0^2n\kappa+1/T'_2$. For large $\chi$, i.e.~strong
measurement, the measurement rate exceeds $\kappa$ leading to a
qubit transition frequency which depends on the instantaneous
value of the cavity photon number and hence an inhomogeneously
broadened gaussian line, see
Fig.~\ref{fig:MeasurementLineWidthCombi}b, whose variance is
simply $\sqrt n$ multiplied by the Stark shift per photon. The
full cross-over from intrinsic lorentzian lineshape with width
$\propto n$ at small $n$ to gaussian lineshape with width $\propto
\sqrt{n}$ at large $n$ as described by Eq.~(4) is in excellent
agreement with the measured dependence of the line width on $n$,
see Fig.~\ref{fig:acStarkShiftLineWidthCombi}b.

In our experiments we have demonstrated that the strong coupling
of a Cooper pair box to a non-resonant microwave field in a
on-chip cavity gives rise to a large qubit dependent shift in the
excitation energy of the resonator. This is used to perform a QND
measurement of the qubit. In the qubit, the ac-Stark effect shifts
the qubit level separation by about one line width per photon and
the back-action of the fluctuations in the field gives rise to a
large broadening of the qubit line.

\begin{acknowledgments}
We would like to thank Michel Devoret for numerous discussions.
This work was supported in part by the National Security Agency
(NSA) and Advanced Research and Development Activity (ARDA) under
Army Research Office (ARO) contract number DAAD19-02-1-0045, the
NSF ITR program under grant number DMR-0325580, the NSF under
grant number DMR-0342157, the David and Lucile Packard Foundation,
the W.~M.~Keck Foundation, and the Natural Science and Engineering
Research Council of Canda (NSERC).
\end{acknowledgments}
 \vspace*{-4ex}


\end{document}